\documentclass[prb,amsmath,amssymb,twocolumn]{revtex4-2}
\usepackage{graphicx}
\usepackage{subfigure}
\usepackage{adjustbox}
\usepackage{bm}
\usepackage{color}
\usepackage{braket}
\usepackage{standalone}
\usepackage{multirow}
\usepackage{tikz}
\usepackage{mathrsfs}
\usepackage{dsfont}
\usepackage[colorlinks,bookmarks=true,citecolor=blue,linkcolor=blue,urlcolor=blue]{hyperref}

\begin{document}

\title{Dissipative preparation of fractional Chern insulators}
\author{Zhao Liu$^1$, Emil J. Bergholtz$^2$, Jan Carl Budich$^3$}
\affiliation{$^1$Zhejiang Institute of Modern Physics, Zhejiang University, Hangzhou 310027, China\\
$^2$Department of Physics, Stockholm University, AlbaNova University Center, 10691 Stockholm, Sweden\\
$^3$Institute of Theoretical Physics, Technische Universit\"at Dresden and W\"urzburg-Dresden Cluster of Excellence ct.qmat, 01062 Dresden, Germany}
\date{\today}

\begin{abstract}
We report on the numerically exact simulation of the dissipative dynamics governed by quantum master equations that feature fractional quantum Hall states as unique steady states. In particular, for the paradigmatic Hofstadter model, we show how Laughlin states can be to good approximation prepared in a dissipative fashion from arbitrary initial states by simply pumping strongly interacting bosons into the lowest Chern band of the corresponding single-particle spectrum. While pure (up to topological degeneracy) steady states are only reached in the low-flux limit or for extended hopping range, we observe a certain robustness regarding the overlap of the steady state with fractional quantum Hall states for experimentally well-controlled flux densities. This may be seen as an encouraging step towards addressing the long-standing challenge of preparing strongly correlated topological phases in quantum simulators.
\end{abstract}
\maketitle

\section{Introduction}
Strongly correlated topological phases that conceptually elude both the independent-particle approximation and a classification in terms of local order parameters exhibit some of the most fascinating phenomena known in nature \cite{Prange1990, Nayak2008, Wen2017}. The influence of dissipation on such systems, while being to some extent inevitable, may be seen as a mixed blessing. On one hand, thermal fluctuations and strong environmental couplings may challenge the desirable topological quantization of observables as well as the coherence of quantum information encoded in topological states. On the other hand, from a viewpoint of state preparation, some form of dissipation is essential to reach topological phases, as is clear from their very definition as equivalence classes under local unitary transformations \cite{Chen2010, Wen2017}. More specifically, local coherent physical processes, mathematically described precisely by such local unitary transformations \cite{Chen2010}, would never allow a system to dynamically enter a topological phase.

For electronic materials, electron-phonon coupling is typically capable of establishing thermal equilibrium with their surroundings, and the practical challenge in preparing a topological state thus amounts to reaching cryostat temperatures below the scale set by the energy gap protecting the targeted topological phase in a given material \cite{Prange1990}. By contrast, a main feature of synthetic materials, e.g., based on ultracold atoms in optical lattices \cite{Bloch2008}, is their high degree of quantum coherence and experimental control over microscopic processes \cite{Bakr2009,Sherson2010,Bloch2012,Jotzu2014,Flaschner2016}. While detrimental effects of dissipation may thereby be strongly contained, a generic analog of the aforementioned cryostat cooling is not naturally present in such quantum simulators \cite{Bloch2008, Langen2015}. Instead, engineered dissipation \cite{Diehl2008,Kraus2008,Verstraete2009,Krauter2011} has been proposed as a means to prepare complex many-body states in a nonequilibrium fashion as steady states of a quantum master equation \cite{Lindblad1976,Breuer2007} from basically arbitrary initial states. This approach may be understood as a tailor-made cooling or entropy reduction protocol for a specific target state rather than a generic thermalization process, and has been discussed for a range of topological states \cite{Diehl2011, Bardyn2013, Budich2015, Goldman2016, Goldstein2019, Shavit2020, Tonielli2020, Bandyopadhyay2020} within the realm of non-interacting topological insulator phases \cite{Hasan2010}. By contrast, there is relatively little corresponding effort \cite{Roy2020,Santos2020} on the engineered dissipation of strongly correlated states.

\begin{figure}
\centerline{\includegraphics[width=\linewidth] {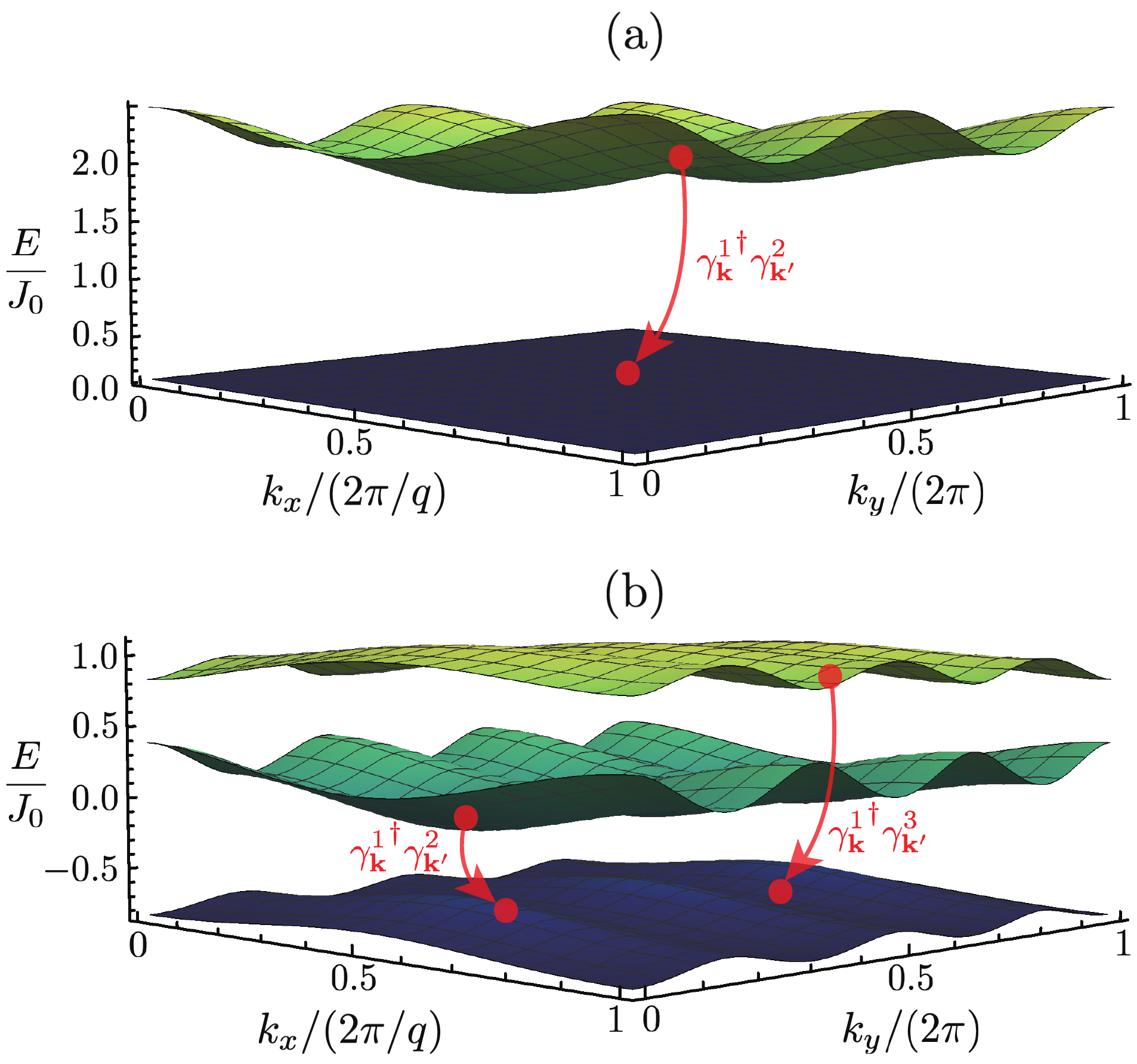}}
\caption{The band structure for (a) the KM model \cite{kapit} with $q=2$ and (b) the Hofstadter model \cite{Hofstadter76} with $q=3$. The red arrows indicate the dissipative pumping of particles from higher bands to the lowest band, as described by the Lindblad operators $L_m$ in Eq.~(\ref{master}).}
\label{fig1}
\end{figure}

Here, turning to strongly correlated dissipative systems, we microscopically simulate the dissipative state preparation of paradigmatic fractional quantum Hall (FQH) states \cite{Tsui1982, Laughlin1983, Prange1990, Bergholtz2013, Parameswaran2013, Halperin2020} in the hard-core boson limit \cite{Kalmeyer1987, YFWang11,kapit} by numerically exact methods. In particular, we construct and study a quantum master equation from which lattice FQH states can be shown to emerge as exact and unique steady states (see Fig.~\ref{fig1} for an illustration). Considering various deviations from this analytically amenable model, we study the robustness of the resulting steady states and their topological properties. While the notion of adiabatic continuity can be readily generalized from Hamiltonian systems to Gaussian states evolving under bilinear Liouvillians \cite{Prosen2010, Eisert2010, Diehl2011}, the absence of such general continuity arguments for our present setting of master equations involving quartic terms serves as a main motivation for our in-depth computational study. 

Notably, for the celebrated Hofstadter model \cite{Hofstadter76} of strongly interacting bosons subject to a magnetic flux \cite{Moeller2009}, we provide numerical evidence that the Laughlin FQH state \cite{Laughlin1983,Kalmeyer1987} can be prepared by a dissipation simply pumping particles to the lowest band of the corresponding noninteracting model. While a pure (up to topological degeneracy) FQH steady state is only reached in the low-flux limit, our data indicate that a satisfactory approximation corresponding to an FQH phase at low temperature may already be reached at experimentally well-studied finite flux \cite{Aidelsburger2013, Goldman2016}. This result is encouraging regarding the long-standing goal of preparing strongly correlated topological phases in synthetic materials. Complementary to our present study, protocols for dissipatively stabilizing FQH states via the quantum Zeno effect \cite{Roncaglia2010}, self-stabilization of initial FQH states \cite{Kapit2014}, and preparing few-body FQH states \cite{Colladay2021} have been reported, and a nonequilibrium topological field theory has been constructed for free Chern insulators from quartic Liouvillians \cite{Tonielli2020}. Furthermore, open systems in an FQH regime have been recently studied \cite{Yoshida2019} employing an effective non-Hermitian Hamiltonian approach \cite{NHbook,NHreview}, as well as considering their stability against quantum jumps \cite{Yoshida2020}.

\section{Model and method}
\subsection{Static Hamiltonian}
We study $N$ on-site interacting bosons in a periodic two-dimensional (2D) square lattice in the $xy$ plane with unit lattice spacing. Each lattice site $j$ is labeled by its position $(x_j,y_j)$, where $x_j=0,1,\cdots,L_x-1$ and $y_j=0,1,\cdots,L_y-1$. The lattice is pierced by a uniform magnetic field, such that the number of flux quanta in an elementary plaquette is a rational number $\phi=p/q$ with coprime integers $p$ and $q$, leading to $q$ magnetic Bloch bands. We choose $q$ sites in the $x$ direction as a magnetic unit cell, so the numbers of unit cells are $N_x=L_x/q$ and $N_y=L_y$ in the $x$ and $y$ directions, respectively. In this paper, we assume $L_x$ to be divisible by $q$ to ensure an integer number of unit cells and focus on $\phi=1/q$.

Based on the scenario described above, we identify desirable target states for our subsequent dissipative state preparation analysis by first considering the static physics governed by the tight-binding Hamiltonian
\begin{eqnarray}
H = -\sum_{j \neq k} t_{jk} a_{j}^{\dagger} a_{k} + U\sum_i n_i(n_i-1),
\label{Hamil}
\end{eqnarray}
where $a_j^\dagger$ ($a_j$) creates (annihilates) a boson on site $j$, $U>0$ is the strength of the onsite repulsion, and $t_{jk}$ is the hopping coefficient between sites $j$ and $k$.

The single-particle physics is determined by $t_{jk}$. If we choose
\begin{eqnarray}
t_{jk}=\sum_{s,t=-\infty}^{+\infty}J(x_j+sL_x,y_j+tL_y;x_k,y_k)e^{-2\pi i s L_x y_j \phi},\nonumber\\
\label{hopping}
\end{eqnarray}
where 
\begin{eqnarray}
J(x_j,y_j;x_k,y_k)=J_0(-1)^{x+y+xy} e^{-\frac{\pi}{2}(1-|\phi|)(x^2+y^2)} \nonumber\\
\times e^{i\pi\phi(x_j+x_k)y}
\end{eqnarray}
with $x=x_j-x_k$ and $y=y_j-y_k$, the setup corresponds to the Kapit-Mueller (KM) model~\cite{kapit} in the Landau gauge. Note that in this case the hopping strength decays exponentially with the hopping range, and this decay is quicker for smaller $\phi$.
At $\phi=1/q$, the lowest band of the KM model has an elegant analytical property: It is spanned by the discretized version of the continuum lowest Landau level (LLL) wavefunctions and is exactly flat carrying Chern number $\mathcal{C}=1$~\cite{kapit}. This analytical similarity to the LLL also allows for the generalization of a number of many-body results found in the continuum LLL. In particular, the ground state of the static Hamiltonian Eq.~(\ref{Hamil}) at filling $\nu\equiv N/(N_xN_y)=1/2$ is the exact Laughlin state discretized to the lattice and possesses zero energy~\cite{kapit}. These exact Laughlin states on the lattice completely reside in the lowest band of the KM model, and their vanishing property (i.e., two bosons cannot be located at the same position) guarantees that the occupation of bosons forming these states is no more than one on each lattice site. These properties survive even when the interaction strength $U\rightarrow+\infty$. In this limit, we can see from Eq.~(\ref{Hamil}) that the exact Laughlin state is the unique state at $\nu=1/2$ that completely resides in the lowest KM band.

Besides the KM model, we will also consider other lattice models with $t_{jk}$ truncated to the nearest-neighbor (NN) and next-nearest-neighbor (NNN) sites. When only the NN hopping exists, we have the usual Hofstadter model~\cite{Hofstadter76}. Note that the capability of realizing the exact $\nu=1/2$ Laughlin state on the KM lattice originates from the analytical similarity between the lowest KM band and the LLL, and is thus in general impossible in other lattice models.

\subsection{Dissipative dynamics}
Our aim is to theoretically model and analyze the preparation of the $\nu=1/2$ bosonic Laughlin phase through dissipative dynamics. Therefore we fix the filling at $\nu=1/2$ throughout this paper. At $t=0$, we prepare the system in a state described by a density matrix $\rho(0)$. Under the assumption of a Markovian process, the dissipative dynamics is governed by the master equation
\begin{eqnarray}
\frac{d\rho}{dt}=-\frac{i}{\hbar}[H,\rho]+\sum_m \kappa_m\left[L_m \rho L_m^\dagger-\frac{1}{2}\left\{L_m^\dagger L_m,\rho\right\}\right],\nonumber\\
\label{master}
\end{eqnarray}
where $\rho\equiv\rho(t)$ is the density matrix of the system at time $t$ and $L_m$'s are the Lindblad operators of strength $\kappa_m$ accounting for dissipative processes within the Born-Markov approximation. 

Motivated by the fact that for the KM model the exact $\nu=1/2$ Laughlin state is the unique state completely residing in the lowest band when $U\rightarrow+\infty$, we choose the dissipative processes that pump bosons from excited bands to the lowest band and meanwhile assume the hard-core limit for bosons. Such processes may, for example, be induced by collisional coupling of the bosons to atoms in a low-temperature Bose-Einstein-condensate (BEC)\cite{Griessner2007}. The generic setting of dissipative pumping to the lowest band will be considered not only for the KM model but also for models with truncated hopping. In this scenario, the Lindblad operator is of the form $L_m\equiv L_{s>1,{\bf k},{\bf k}'}={\gamma_{\bf k}^1}^\dagger \gamma_{{\bf k}'}^{s>1}$, which moves a particle from an excited band $s>1$ to the lowest band and meanwhile scatters its crystal momentum from ${\bf k}'$ to ${\bf k}$ (see Fig.~\ref{fig1} for an illustration). Here, ${\gamma_{\bf k}^s}^\dagger$ creates a particle with momentum ${\bf k}$ in band $s$, and ${\bf k}$ is restricted in the first Brillouin zone (1BZ). There are $(q-1)(N_xN_y)^2$ Lindblad operators, where $q-1$ is the number of higher bands and $N_xN_y$ is the number of ${\bf k}$ points in the 1BZ. For simplicity, we set $\kappa_m=\kappa$ for all Lindblad operators.

To impose the hard-core boson condition, we Fourier-transform these Lindblad operators and the master equation Eq.~(\ref{master}) from momentum space to real space (see details in Appendix~\ref{appA}). Then we construct the real-space Fock basis in which the number of bosons on each lattice site is no more than $1$. The Lindblad operators, the many-body Hamiltonian, and the density matrix are all represented as matrices under this basis. In the hardcore-boson limit, only the hopping term exists in the many-body Hamiltonian Eq.~(\ref{Hamil}), while the hardcore constraint is encoded in the restricted Hilbert space as compared with bosons with finite interaction.

When simulating the dynamics, we divide time into many steps of small interval $\Delta t$ and discretize Eq.~(\ref{master}). Given the state $\rho(t)$ at time $t$, the state at time $t+\Delta t$ is evaluated by
\begin{eqnarray}
\rho(t+\Delta t)=\rho(t)&-&\frac{i}{\hbar}[H,\rho(t)]\Delta t
+\sum_m \kappa_m[L_m \rho(t) L_m^\dagger\nonumber\\
&-&\frac{1}{2}\left\{L_m^\dagger L_m,\rho(t)\right\}]\Delta t.
\label{discrete}
\end{eqnarray}
For our specific setting, our examination shows that it is typically sufficient to choose $\kappa\Delta t=0.1$ and further reduction of $\Delta t$ gives very similar results. Within the limit of our numerical resources, we can only deal with at most four bosons at a numerically exact level. In this context, it is worth noting that the dimension of the considered vector space in which the density matrix is defined grows quadratically with the Hilbert space dimension of pure state vectors.

We will focus on the pure dissipation case, i.e., we neglect the effect of the $-\frac{i}{\hbar}[H,\rho]$ term. This is reasonable when the Hamiltonian and the Lindblad operators come from the same lattice model such that they are compatible with each other, or in a scenario where the target system may be viewed as a (to good approximation) flat band of a Hamiltonian. We have checked that bringing the $-\frac{i}{\hbar}[H,\rho]$ term with a compatible Hamiltonian back leads to even quantitatively quite similar results and identical conclusions.

For a specific lattice model (KM or truncated model), we characterize the dynamics of $\rho(t)$ by two quantities. First, its overlap 
\begin{eqnarray}
\label{ot}
\mathcal{O}(t)\equiv\sum_{i=1}^2\langle\Psi_i|\rho(t)|\Psi_i\rangle
\end{eqnarray}
with the twofold degenerate topologically ordered ground states $|\Psi_i\rangle$ of the corresponding static Hamiltonian on the torus, which are numerically obtained by diagonalizing the Hamiltonian shown in Eq.~(\ref{Hamil}). For the KM model in the hardcore limit, the Hamiltonian ground states are simply the exact Laughlin states, and otherwise FQH states in the same topological phase as the $\nu=1/2$ Laughlin state. Second, its weight \begin{eqnarray}
\label{wt}
\mathcal{W}(t)\equiv\frac{1}{N}\sum_{{\bf k}}{\rm Tr}[\rho(t){\gamma_{\bf k}^1}^\dagger\gamma_{\bf k}^1]
\end{eqnarray}
in the lowest band. The calculation of $\mathcal{O}(t)$ is straightforward because both $|\Psi_i\rangle$ and $\rho(t)$ are expressed in the real-space basis corresponding to the square lattice geometry. When evaluating $\mathcal{W}(t)$, we need to transform ${\gamma_{\bf k}^1}^\dagger\gamma_{\bf k}^1$ to the real space first (see details in Appendix~\ref{appA}).

\section{Dynamics of the KM model}
As for the KM model the exact Laughlin state is the unique state at $\nu=1/2$ that completely resides in the lowest band, we expect that our state-preparation protocol will select the exact $\nu=1/2$ Laughlin state as the steady state in this case for any initial state with the right particle number. To numerically confirm this, we consider a pure initial state, which can be either a product state randomly chosen from the Fock basis, or a random superposition of all basis states. We find similar results in both cases, so we demonstrate the dynamics starting from a product state in the following.

\begin{figure}
\centerline{\includegraphics[width=\linewidth] {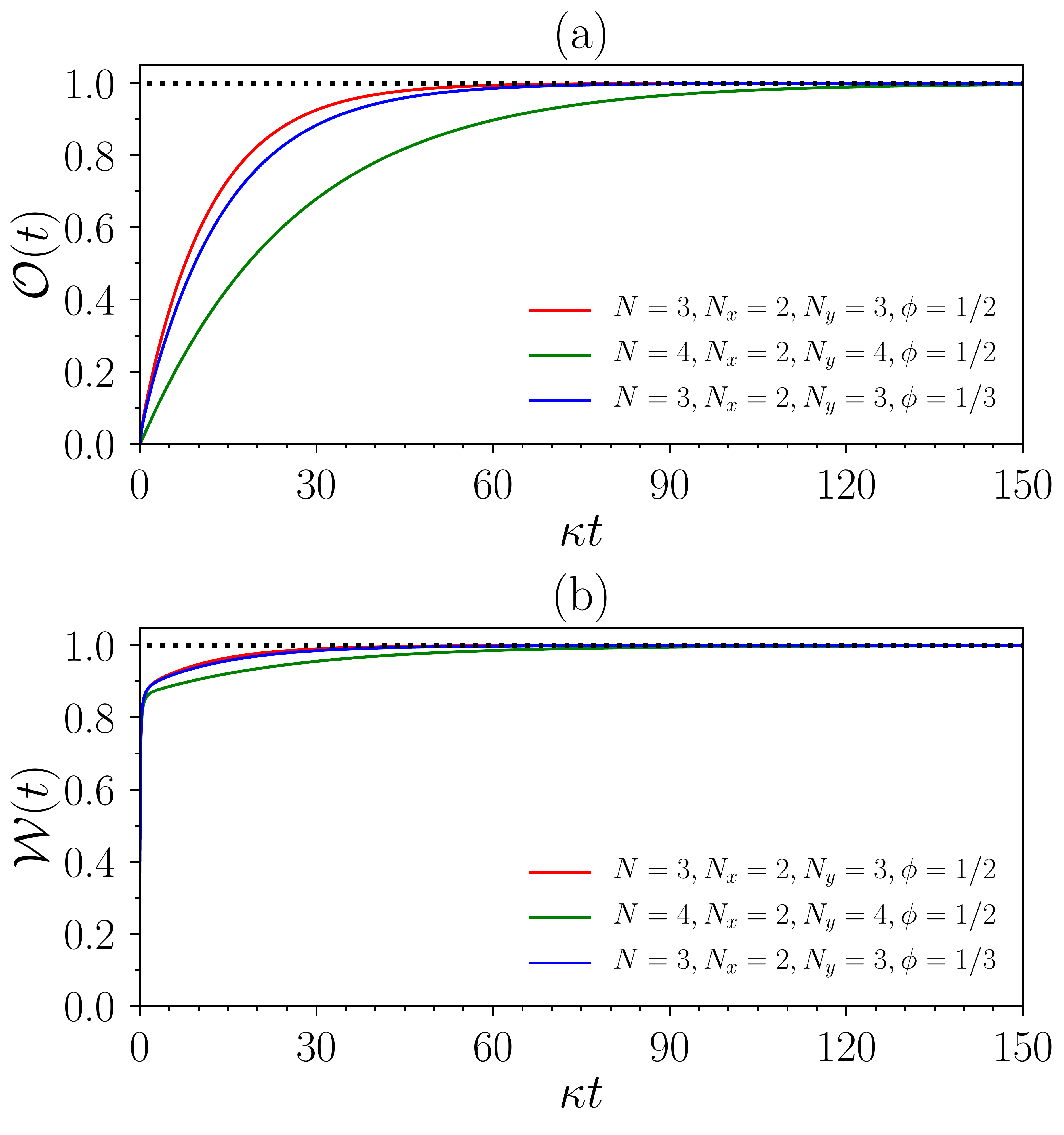}}
\caption{The dynamics Eq.~(\ref{master}) for the KM model. We characterize the dynamics by (a) the overlap Eq.~(\ref{ot}) with the exact $\nu=1/2$ Laughlin states and (b) the weight Eq.~(\ref{wt}) in the lowest KM band as functions of $\kappa t$. The dotted lines indicate $\mathcal{O}(t)=1$ and $\mathcal{W}(t)=1$ in (a) and (b), respectively.}
\label{KM}
\end{figure}

The dynamics at flux densities $\phi=1/2$ and $\phi=1/3$ is shown in Fig.~\ref{KM}. For all system sizes that we have studied, both $\mathcal{O}(t)$ and $\mathcal{W}(t)$ tend to $100\%$ in the long-time limit. This unambiguously indicates that for the KM model the dissipation governed by particles' jumping from higher bands to the lowest band indeed drives the system to a steady state which is completely located in the subspace of exact Laughlin states, i.e., a unique steady state up to topological degeneracy. The purity of $\rho(t)$, defined as ${\rm Tr}\rho^2(t)$, is $0.5$ in the long-time limit, meaning that the steady state is a mixed state with $50\%$ probability on each of the two exact Laughlin states on the torus. 

\section{Beyond the KM model}
Having confirmed that we can perfectly prepare the exact $\nu=1/2$ Laughlin state in the KM model by the dissipative dynamics, we find it interesting to investigate to what extent the strategy of pumping all particles to the lowest band of a single-particle picture still works when deviating from the KM model. The answer to this question is unclear due to the absence of analytical similarity to the LLL in generic lattice models. In particular, it is not obvious whether a state close to the exact $\nu=1/2$ Laughlin state exists completely in the lowest band of the pertinent model in the hard-core limit. Furthermore, in general there is no strict adiabatic continuity in the steady state of non-Gaussian Liouvillians towards changes in the Lindblad operators.

\begin{figure}
\centerline{\includegraphics[width=\linewidth] {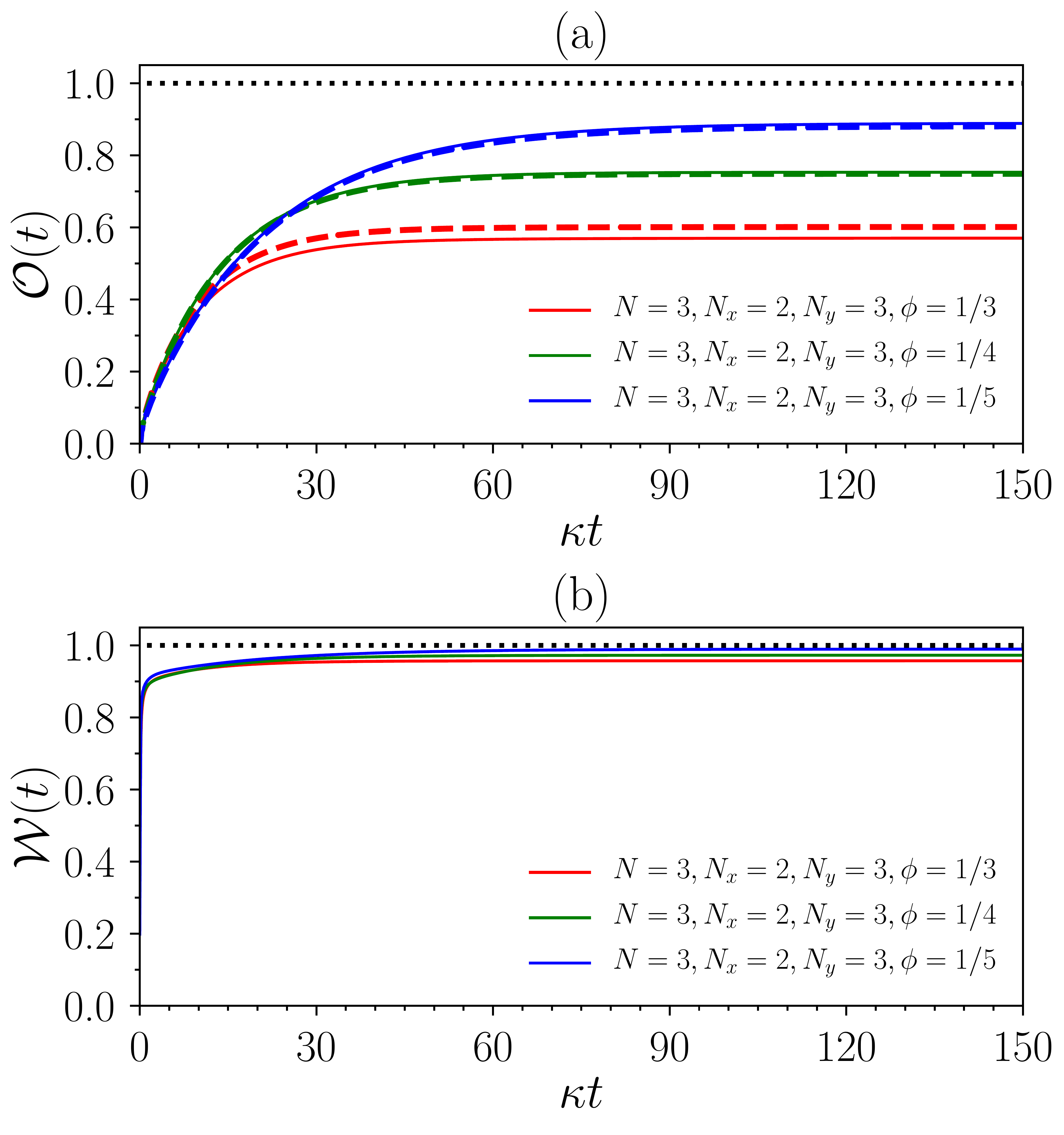}}
\caption{The dynamics Eq.~(\ref{master}) for the Hofstadter model. We characterize the dynamics by (a) the overlaps Eq.~(\ref{ot}) with the exact $\nu=1/2$ Laughlin states (dashed curves) and the Hofstadter ground states (solid curves) and (b) the weight Eq.~(\ref{wt}) in the lowest Hofstadter band as functions of $\kappa t$. The dotted lines indicate $\mathcal{O}(t)=1$ and $\mathcal{W}(t)=1$ in (a) and (b), respectively.}
\label{Hof}
\end{figure}

To be concrete, let us first focus on the truncated model with only the NN hopping, i.e., the Hofstadter model. Similar to the KM case, we still choose a product state as the initial state; however, the Lindblad operators should now be defined using the Bloch states of the Hofstadter model, corresponding to particles' jumping from higher Hofstadter bands to the lowest Hofstadter band. As a Dirac-type band touching exists for the Hofstadter model at $\phi=1/2$, thus rendering the corresponding Hamiltonian critical, we consider $q\geq 3$ in the following, such that a finite gap above the lowest band occurs in the corresponding Hamiltonian model and the static Hofstadter ground state at $\nu=1/2$ has a high overlap with the exact Laughlin state.  
The evolution of $\mathcal{O}(t)$ and $\mathcal{W}(t)$ for three bosons on the $2\times 3$ Hofstadter lattice with $\phi=1/3,1/4$ and $1/5$ is shown in Fig.~\ref{Hof}. In these cases, both $\mathcal{O}(t)$ and $\mathcal{W}(t)$ saturate at long time, suggesting that the Hofstadter system still reaches a steady state.
However, the overlaps between the steady state and the static Hofstadter ground state only reach $57.0, 75.3$ and $88.8\%$ for $\phi=1/3,1/4$, and $1/5$, respectively [Fig.~\ref{Hof}(a)], which means that the dissipative dynamics does not drive the system to the subspace of static FQH ground states. The corresponding overlaps with the exact Laughlin state are $60.1, 74.9$ and $87.7\%$ for $\phi=1/3,1/4$, and $1/5$, respectively, which are much lower compared with the KM case especially at high flux density $\phi=1/3$.

\begin{figure}
\centerline{\includegraphics[width=\linewidth] {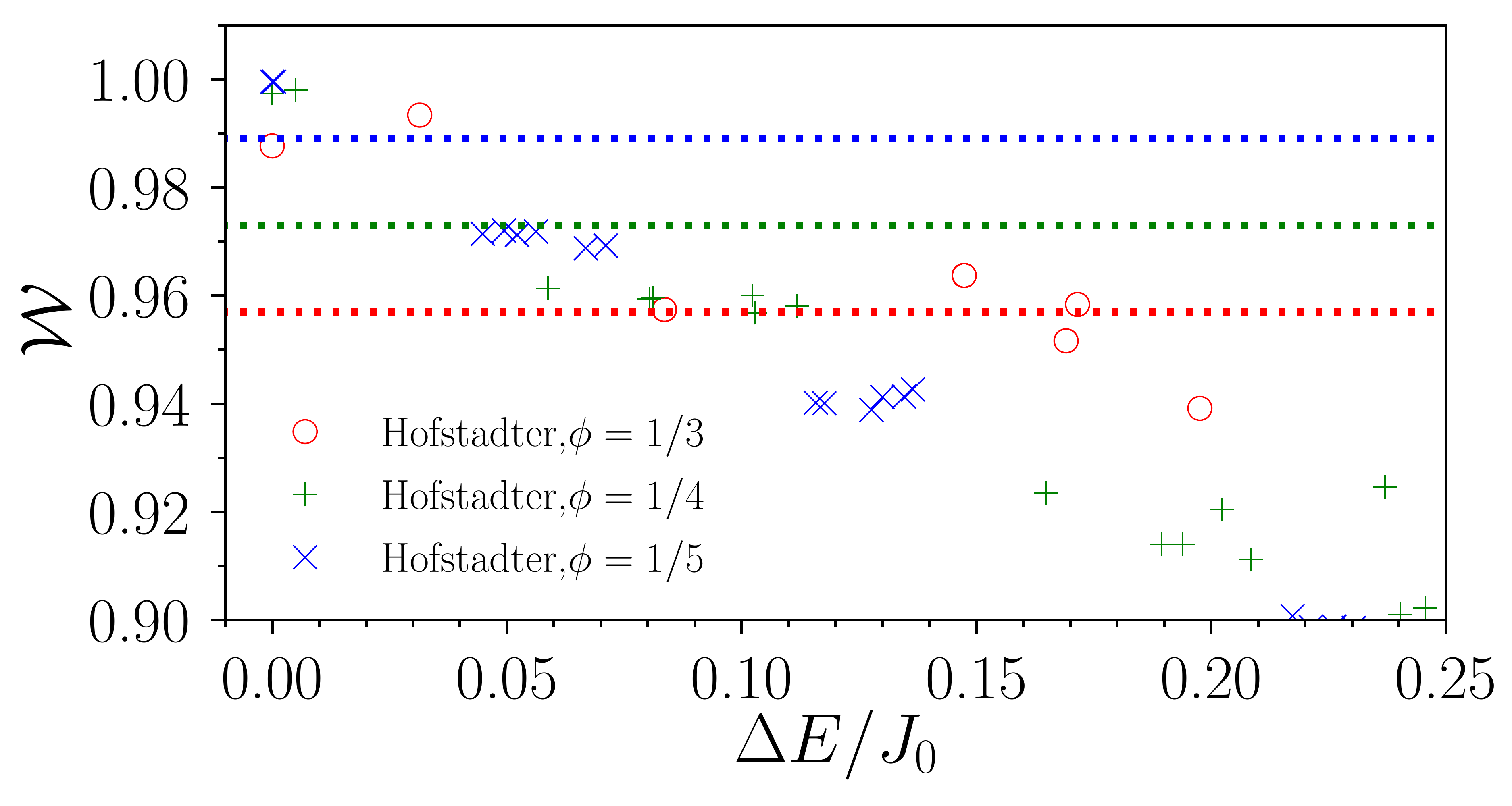}}
\caption{The weight $\mathcal{W}$ [Eq.~(\ref{wt})] of the many-body eigenstate of the static Hamiltonian Eq.~(\ref{Hamil}) in the lowest Hofstadter band for $N=3, N_x=2, N_y=3$ at $\phi=1/3,1/4$, and $1/5$. Each eigenstate is labeled by its excitation energy $\Delta E$ with respect to the ground state. The dotted lines indicate the steady state's weight in the lowest Hofstadter band [shown in Fig.~\ref{Hof}(b)] for the corresponding system sizes.}
\label{bw}
\end{figure}

The worse performance of preparing the Laughlin phase in the Hofstadter model (especially at high flux density) is due to the absence of a state thereof at $\nu=1/2$ that is close to the exact Laughlin state and completely resides in the lowest Hofstadter band in the hard-core limit. In Fig.~\ref{bw}, we plot the weights in the lowest Hofstadter band for all eigenstates of the Hofstadter model at $\nu=1/2$ in the hard-core limit. While the ground states have higher weights than excited states, they do not completely reside in the lowest Hofstadter band, which is more obvious for larger $\phi$ (Fig.~\ref{bw}). Therefore the strategy of pumping particles to the lowest Hofstadter band does not necessarily drive the system to the ground-state subspace of the Hofstadter model, which has a high overlap with the exact Laughlin state. Instead, some excited eigenstates far from the exact Laughlin states may be involved, leading to the relatively low overlap between the steady state and the exact Laughlin state. This is also supported by our numerical observation that the weight of the steady state in the lowest Hofstadter band is lower than that of the static ground state ($95.7$ versus $99\%$ for three bosons on the $2\times 3$ Hofstadter lattice at $\phi=1/3$), suggesting the contribution from static excited eigenstates to the steady state [Fig.~\ref{Hof}(b) and Fig.~\ref{bw}]. By contrast, for the KM model at $\nu=1/2$, all particles can jump to the lowest KM band and meanwhile satisfy the hard-core condition by forming the exact Laughlin state, such that our dissipation mechanism naturally selects the Laughlin state as the steady state. 

\begin{figure}
\centerline{\includegraphics[width=\linewidth] {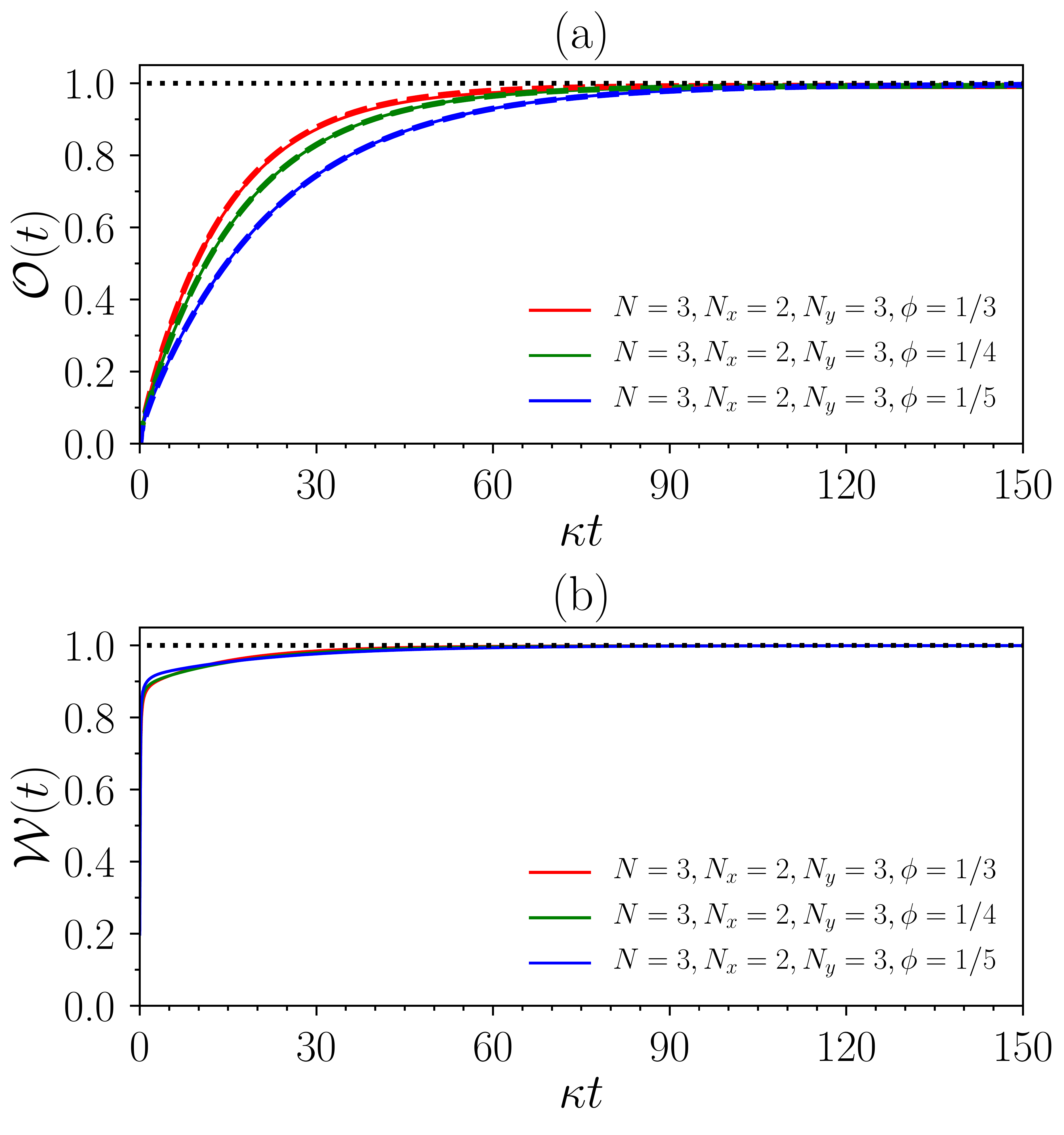}}
\caption{The dynamics Eq.~(\ref{master}) for the Hofstadter model with additional NNN hopping. We characterize the dynamics by (a) the overlaps Eq.~(\ref{ot}) with the $\nu=1/2$ exact Laughlin states (dashed curves) and the ground states of the model (solid curves) and (b) the weight Eq.~(\ref{wt}) in the lowest band as functions of $\kappa t$. The discrepancy between the solid curve and the dashed curve is invisible in (a) for $\phi=1/4$ and $\phi=1/5$. The dotted lines indicate $\mathcal{O}(t)=1$ and $\mathcal{W}(t)=1$ in (a) and (b), respectively.}
\label{truncate2}
\end{figure}

There are two ways to gain better preparation of the Laughlin state in the Hofstadter model. One is to decrease the flux density, because in this case the Hofstadter model is closer to the KM model due to the fast decaying of longer-range KM hopping. Indeed, for three bosons on the $2\times 3$ Hofstadter lattice, the overlap between the steady state and the static Hofstadter ground state increases from $57.0$ to $88.8\%$ (the overlap with the exact Laughlin state increases from $60.1$ to $87.7\%$) when $\phi$ is reduced from $1/3$ to $1/5$ [Fig.~\ref{Hof}(a)], and the weight of the steady state in the lowest Hofstadter band increases from $95.7$ to $98.9\%$ [Fig.~\ref{Hof}(b)]. 

Alternatively, we can add longer hopping in the Hofstadter model. Remarkably, we observe a significant improvement of the results once we add the next-nearest-neighbor (NNN) hopping. In Fig.~\ref{truncate2}, we show the results obtained by keeping the NN and NNN hopping in the original KM model for three bosons on the $2\times 3$ lattice. Compared with the data in Fig.~\ref{Hof}, the long-time overlap with the static ground state of the model dramatically grows from $57.0$ to $99.4\%$ (the overlap with the exact Laughlin state dramatically grows from $60$ to $99.4\%$) for $\phi=1/3$ [Fig.~\ref{truncate2}(a)], accompanied by an increasing of the lowest-band weight from $95.7$ to $99.9\%$ [Fig.~\ref{truncate2}(b)]. A similar improvement can also be seen for $\phi=1/4$ (Fig.~\ref{truncate2}). 

In summary, the proposed state-preparation protocol of band pumping still works excellently at low flux densities for realistic models with only the NN hopping. Further improvement can be obtained by adding slightly longer range hopping. 

\section{Concluding discussion}
Motivated by the important open challenge of preparing strongly correlated topological phases in quantum simulators, we have numerically studied the Liouvillian dynamics of microscopic lattice models towards fractional quantum Hall steady states. For the experimentally well-studied Hofstadter model in the low-flux limit, as well as for the Kapit-Mueller model characterized by an extended hopping range, bilinear Lindblad jump operators (leading to a quartic Liouvillian) that yield a pure FQH steady state (up to topological degeneracy) can readily be constructed guided by the physical picture of pumping particles into the lowest Chern band. Beyond these exact limits, when generalizing the aforementioned band pumping picture to experimentally studied finite flux densities, we find an encouraging robustness in the sense of moderate and continuous reduction of the steady state overlaps with FQH model states in a wide parameter range. On a more general note, compared with adiabatic preparation protocols of topological phases the paradigm of dissipative state preparation does not suffer from an unavoidable critical slow-down when approaching the thermodynamic limit.  

While this paper provides a first fully microscopic study on the dissipative preparation of FQH states from arbitrary initial states, several issues clearly remain interesting subjects of future research. First, the pumping rates of the Lindblad jump operators are assumed constant for simplicity. Estimating the momentum dependence of such rates in an experimentally realistic setting of cold atoms with collisional coupling to a BEC is one direction devising a feasible protocol based on our general analysis. Second, in the interest of computational feasibility, our numerically exact study is based on systems of hard-core bosons, which naturally raises the question of the influence of finite contact interactions. To address this issue qualitatively, we argue that taking the hard-core limit is by no means expected to be pathological in the present context. Specifically, an experimental protocol could start from a Mott state without initial double occupations. The dynamical creation of such energetically very costly double occupations (that would kick the system out of the considered hard-core Hilbert space) may then be practically prohibited already at large but finite interaction strength. This is because only energetically resonant processes (in the combined system and bath setting) contribute to the Lindblad operators due to the underlying Born-Markov approximation.  
Finally, although generalizations of the KM model to non-Abelian and higher-Chern-number states provide a natural starting point \cite{jorg}, a similarly simple mechanism for the preparation of more complex FQH states than Laughlin phases remains to be identified.\\

\acknowledgements
Z.L. is supported by the National Natural Science Foundation of China through Grant No.~11974014. E.J.B. is supported by the Swedish Research Council (VR) and the Wallenberg Academy Fellows program as well as the project Dynamic Quantum Matter of the Knut and Alice Wallenberg Foundation. J.C.B. is supported by the German Research Foundation (DFG) through Collaborative Research
Centre SFB 1143 (Project No.~247310070), the Cluster of Excellence ct.qmat (Project No.~390858490), and
DFG Project No.~419241108.

\appendix
\section{Lindblad operators and the master equation in real space}
\label{appA}
Using the Fourier transform 
\begin{eqnarray}
a_{{\bf k},\alpha}^\dagger=\frac{1}{\sqrt{N_x N_y}}\sum_{m=1}^{N_x N_y} e^{i{\bf k}\cdot {\bf R}_m} a_{m,\alpha}^\dagger
\end{eqnarray}
and the relation 
\begin{eqnarray}
{\gamma_{\bf k}^s}^\dagger =\sum_{\alpha=1}^{q}v_{\alpha}^s({\bf k}) a_{{\bf k},\alpha}^\dagger,
\end{eqnarray}
we can express our Lindblad operators in real space as
\begin{eqnarray}
{\gamma_{\bf k}^1}^\dagger \gamma_{{\bf k}'}^{s}=\frac{1}{N_x N_y} \sum_{m,n=1}^{N_x N_y}\sum_{\alpha,\beta=1}^{q}v_{\alpha}^1({\bf k}) {v_{\beta}^{s}}^*({\bf k}') e^{i{\bf k}\cdot {\bf R}_m}\nonumber\\
\times e^{-i{\bf k}'\cdot {\bf R}_n} a_{m,\alpha}^\dagger a_{n,\beta},
\end{eqnarray}
where ${\bf R}_m$ is the position of the $m$th unit cell, $a_{m,\alpha}^\dagger$ creates a boson in the $\alpha$th site of the $m$th unit cell, and $[v_1^s({\bf k}),\cdots,v_q^s({\bf k})]$ is the eigenvector of band $s$.

The master equation with pure dissipation can then be written in real space as 
\begin{eqnarray}
\frac{d\rho}{dt}=\sum_{{\bf k},{\bf k}'\in{\rm 1BZ}}\sum_{s=2}^{q}\left[{\gamma_{\bf k}^1}^\dagger \gamma_{{\bf k}'}^{s}\rho{\gamma_{{\bf k}'}^s}^\dagger \gamma_{{\bf k}}^{1}-\frac{1}{2}\left\{{\gamma_{{\bf k}'}^s}^\dagger \gamma_{{\bf k}}^{1}{\gamma_{\bf k}^1}^\dagger \gamma_{{\bf k}'}^{s},\rho\right\}\right]\nonumber\\
\label{eq5}
\end{eqnarray}
with
\begin{widetext}
\begin{eqnarray}
{\gamma_{\bf k}^1}^\dagger \gamma_{{\bf k}'}^{s}\rho{\gamma_{{\bf k}'}^s}^\dagger \gamma_{{\bf k}}^{1}
&=&\frac{1}{(N_x N_y)^2}\sum_{i,j=1}^{D}\rho_{i,j}\sum_{m,n=1}^{N_xN_y}\sum_{\alpha,\beta=1}^q\sum_{m',n'=1}^{N_xN_y}\sum_{\alpha',\beta'=1}^q\Big[
v_{\alpha}^1({\bf k}) {v_{\beta}^{s}}^*({\bf k}') {v_{\alpha'}^1}^*({\bf k}) {v_{\beta'}^{s}}({\bf k}')\nonumber\\ &&\times e^{i{\bf k}\cdot ({\bf R}_m-{\bf R}_m')}e^{-i{\bf k}'\cdot ({\bf R}_n-{\bf R}_n')} \Big] a_{m,\alpha}^\dagger a_{n,\beta} |i\rangle\langle j| a_{n',\beta'}^\dagger a_{m',\alpha'},\nonumber\\
-\frac{1}{2}\left\{{\gamma_{{\bf k}'}^s}^\dagger \gamma_{{\bf k}}^{1}{\gamma_{\bf k}^1}^\dagger \gamma_{{\bf k}'}^{s},\rho\right\}
&=&-\frac{1}{2(N_x N_y)^2}\sum_{i,j=1}^{D}\rho_{i,j}\sum_{m,n=1}^{N_xN_y}\sum_{\alpha,\beta=1}^q\sum_{m',n'=1}^{N_xN_y}\sum_{\alpha',\beta'=1}^q\Big[
v_{\alpha}^1({\bf k}) {v_{\beta}^{s}}^*({\bf k}') {v_{\alpha'}^1}^*({\bf k}) {v_{\beta'}^{s}}({\bf k}')\nonumber\\ &&\times e^{i{\bf k}\cdot ({\bf R}_m-{\bf R}_m')}e^{-i{\bf k}'\cdot ({\bf R}_n-{\bf R}_n')} \Big] \left(a_{n',\beta'}^\dagger a_{m',\alpha'} a_{m,\alpha}^\dagger a_{n,\beta} |i\rangle\langle j| 
+|i\rangle\langle j| a_{n',\beta'}^\dagger a_{m',\alpha'} a_{m,\alpha}^\dagger a_{n,\beta}\right),\nonumber\\
\label{eq6}
\end{eqnarray}
\end{widetext}
where $|i\rangle$ is the Fock basis state of the many-body Hilbert space of dimension $D$ and $\rho_{i,j}$ is the matrix element under this basis. Once we express the master equation in real space as Eqs.~(\ref{eq5}) and (\ref{eq6}), we can easily impose the hardcore constraint by working in the Fock basis where the boson occupation per site is either $0$ or $1$. The weight in the lowest band can be evaluated by 
\begin{widetext}
\begin{eqnarray}
\mathcal{W}=\frac{1}{N}{\rm Tr}\left(\rho\sum_{{\bf k}\in{\rm 1BZ}}{\gamma_{\bf k}^1}^\dagger \gamma_{{\bf k}}^{1}\right)
=\frac{1}{NN_x N_y} \sum_{i,j=1}^D\rho_{i,j}\sum_{m,n=1}^{N_xN_y}\sum_{\alpha,\beta=1}^q\Big[\sum_{{\bf k}\in{\rm 1BZ}}v_{\alpha}^1({\bf k}) {v_{\beta}^{1}}^*({\bf k}) e^{i{\bf k}\cdot ({\bf R}_m-{\bf R}_n)}\Big] \langle j|a_{m,\alpha}^\dagger a_{n,\beta}|i\rangle.
\end{eqnarray}
\end{widetext}

\end{document}